\documentclass[conference]{IEEEtran}
\IEEEoverridecommandlockouts
\usepackage{cite}
\usepackage{amsmath,amssymb,amsfonts}
\usepackage{algorithmic}
\usepackage{graphicx}
\usepackage{textcomp}
\usepackage{xcolor}
\usepackage[english, brazil]{babel}
\usepackage{orcidlink}
\def\BibTeX{{\rm B\kern-.05em{\sc i\kern-.025em b}\kern-.08em
    T\kern-.1667em\lower.7ex\hbox{E}\kern-.125emX}}
\begin{document}

\title{DeLIAP e DeLIAJ:
Interfaces de biblioteca de dependabilidade para Python e Julia\\
}

\author{\IEEEauthorblockN{Marcos Irigoyen\orcidlink{0009-0003-8370-8636}}
\IEEEauthorblockA{\textit{Dep. de Engenharia de Computação e Automação} \\
\textit{Universidade Federal do Rio Grande do Norte}\\
Natal, Brasil \\
marcos.irigoyen.706@ufrn.edu.br}
\and
\IEEEauthorblockN{Carla Santana\orcidlink{0000-0003-3328-0056}}
\IEEEauthorblockA{\textit{Lab. de Arquiteturas Paralelas para Processamento de Sinais} \\
\textit{Universidade Federal do Rio Grande do Norte}\\
Natal, Brasil \\
carla.santana.058@ufrn.edu.br}
\and
\IEEEauthorblockN{Ramon C. F. Araújo\orcidlink{0000-0002-3786-801X}}
\IEEEauthorblockA{\textit{Dep. de Física Teórica e Experimental} \\
\textit{Universidade Federal do Rio Grande do Norte}\\
Natal, Brasil \\
ramon@fisica.ufrn.br}
\and
\IEEEauthorblockN{Samuel Xavier-de-Souza\orcidlink{0000-0001-8747-4580}}
\IEEEauthorblockA{\textit{Dep. de Engenharia de Computação e Automação} \\
\textit{Universidade Federal do Rio Grande do Norte}\\
Natal, Brasil \\
samuel@dca.ufrn.br}
}
\maketitle

\begin{abstract}

O constante crescimento de complexidade computacional de aplicações de processamento de alto desempenho é, comumente, compensado com um escalonamento horizontal dos sistemas computacionais. Colateralmente, cada nó incluído é um potencial ponto de falha para programas paralelos, aumentando a necessidade de implementações de técnicas de tolerância a falhas, sobretudo a nível do programa. Sob tais condições, a biblioteca de tolerância a falhas DeLIA foi desenvolvida em C/C++ com capacidades de detecção de erro e recuperação de falhas. Propomos, então, estender as capacidades da DeLIA para Python e Julia através das interfaces DeLIAP e DeLIAJ a fim de diminuir a barreira de entrada para implementação de soluções de tolerância a falhas nestas linguagens, que carecem de alternativas à biblioteca. Para validar a eficácia dessas interfaces, é analisada sua aplicação no método de inversão de onda completa da forma de onda 4D em Julia, verificando quantitativamente o custo computacional adicionado por comparações entre o tempo de execução, enquanto que, para tratar da aplicabilidade, são apresentados exemplos e relato da implementação. O custo computacional adicionado refletiu em um \textit{overhead} mediano de 1,4\%, enquanto que limitações no módulo para paralelismo usado originalmente para a aplicação inviabilizava o uso de salvamento de dados locais à aplicação.

\textit{Palavras-chave}—Tolerância a falhas, Dependabilidade, PAD, Checkpointing, Julia, Python
\end{abstract}


\section{Introdução}
A tanto tempo quanto temos computadores também temos problemas aos quais aplicamo-los para sua resolução. Estes problemas variam em sua complexidade, requerendo, consequentemente, diferentes quantias de esforço computacional, onde alguns problemas são custosos demais para serem resolvidos em tempo hábil por um único computador comum. Para lidar com estes desafios são desenvolvidos sistemas computacionais especiais, comumente exigindo programas dedicados para aproveitar adequadamente a capacidade desses sistemas \cite{hpc_introduction_hager}. 

Processamento de Alto Desempenho (PAD), então, pode ser generalizado como a área de design, desenvolvimento e implementação de sistemas computacionais, bem como as ferramentas, técnicas e aplicações construídas em volta deles, para resolução de problemas computacionalmente custosos.

A crescente complexidade de novos problemas promove a necessidade de cada vez mais recursos computacionais. Hoje em dia, sistemas de PAD são sistemas distribuídos, evoluindo não somente no desempenho individual de cada máquina, como também escalonando horizontalmente, adicionando nós computacionais para atender a demanda dos programas paralelos \cite{PP_pra_hpc_calebe}.

Esta natureza distribuída chama, por si só, por precauções no desenvolvimento e execução de tais programas, seja para assegurar-se da corretude, operacionalidade e disponibilidade da aplicação, ou minimizar possíveis perdas. Cada componente discreto em um sistema é suscetível a falhas, podendo causar disrupções na sua ocorrência, independentemente da origem da falha. Diante disso, torna-se recomendável a capacitação de sistemas para lidar com falhas evitando interrupções de execução ou perdas totais de dados, ou seja, que os sistemas sejam tolerantes a falhas.

Aplicações de PAD são especialmente frágeis a falhas, dado seu inerentemente maior custo computacional e energético. O uso massivo de nós computacionais em um programa paralelo aumenta o risco de falhas, pois cada nó participante constitui-se um ponto único de falha, capaz de interromper toda a execução em aplicações sem as devidas precauções. Torna-se imperativo, então, o uso de técnicas de tolerância a falhas no contexto de PAD.

Para facilitar o emprego de tolerância a falhas, a biblioteca de confiabilidade para aplicações iterativas DeLIA (do inglês \textit{Dependability Library for Iterative Applications}, desenvolvida na UFRN) foi proposta com capacidades de detecção de interrupção e recuperação de falhas \cite{DeLIA_artigo}. Aqui, propõe-se estender as capacidades da DeLIA para Python (DeLIAP) e Julia (DeLIAJ), linguagens bem estabelecidas em computação científica em geral que, para além da maior facilidade de uso e prototipagem, dispõem de menos alternativas para essas tarefas \cite{2020SciPy-NMeth}\cite{julia_origem_bezanson}.

Para validar a eficácia da solução proposta, aplicamos a interface em Julia em uma base de códigos no método de inversão de onda completa da forma de onda 4D, utilizada internamente no grupo de pesquisa do projeto "Novas metodologias
computacionalmente escaláveis para sísmica 4D orientado ao alvo em reservatórios do
pré-sal" \cite{fwi_residuo_estimado}. Foram feitas análises do custo computacional introduzido pela inclusão da biblioteca através da comparação entre medidas de tempo de execução de uma versão com a DeLIAJ e outra sem a DeLIAJ.

Após detalhamento da base teórica da aplicação e o que se tem desenvolvido e usado atualmente para implementar tolerância a falhas, separamos a metodologia para abordar tanto como foi feito os experimentos de custo computacional quanto a implementação das interfaces nas duas linguagens. Os resultados e sua discussão ocupam a mesma seção, logo em seguida, subdivididos na análise do tempo e no relato de implementação.

\section{Embasamento Teórico}

O crescimento na complexidade de projetos tanto de hardware quanto de software tornam inevitavelmente maior a chance de componentes de qualquer natureza apresentarem um comportamento indesejado. Para conceitualização da ocorrência destes comportamentos são usados os termos \textit{fault}, \textit{error} e \textit{failure}, doravante traduzidos a falha, erro e defeito \cite{ft_terminology}.

Falha é definido como a causa do erro, podendo estar presente em um sistema sem que este tenha sido afetado ainda — como um setor de memória danificado ou erros de implementação em uma seção de código. Uma falha é dita ativada quando sua presença leva o sistema a um estado incorreto, chamado de erro; que pode ser interpretado como um estado intermediário entre a falha e o defeito, onde a continuação da execução sem correção desse estado pode levar a um defeito \cite{conceptual_ft_heimer}\cite{weber_ft_conceitos_e_ex}. O defeito, enfim, é o comportamento desviante do pretendido, como, para aplicações em PAD, a interrupção de um nó computacional, cálculo de valores absurdos e atraso significante na execução.

Ainda nesta definição, falhas podem ser classificadas por temporalidade — entre falhas permanentes, intermitentes ou transientes —, localidade — em qual componente, discreto ou não, a falha ocorre — e efeito — o tipo do comportamento indesejado \cite{conceptual_ft_heimer}. Para este caso, entretanto, vale regredir aos sistemas de modelos de falha implementados para sistemas distribuídos em específico: \textit{fail-stop} (ou \textit{fail-stutter}) e falhas de consistência interativa \cite{failstop}\cite{failstutter}\cite{byzantine_fault}.

O modelo \textit{fail-stop} simplifica o tratamento de falhas através do interrompimento direto da execução da componente onde se localiza a falha pela própria componente, evitando a interação com o resto do sistema, que detecta a falha na componente independentemente \cite{failstop}.

Falhas de consistência interativa, por sua vez, tratam que as componentes defeituosas possam continuar operando no sistema, sem que se tenha noção da ocorrência de falhas e considerando a possibilidade de coordenação entre componentes defeituosas \cite{survey_fault_tolerance_treaster}. Este modelo representa tipos de falhas menos triviais de se tratar, sendo manifestado durante a comunicação entre componentes e, para PAD, dependendo de soluções para verificação específicas à aplicação ou, às custas de desempenho, através de técnicas de redundância computacional \cite{fault_tolerance_techniques_dongarra}.

A tolerância a falhas, então, designa-se à capacidade de um sistema continuar sua operação adequada apesar da ocorrência de falhas. Seguindo às concepções mais gerais, a tolerância a falhas pode ser dividida em detecção de erros, confinamento, recuperação de erros e tratamento da falha \cite{weber_ft_conceitos_e_ex}. Detecção se resumindo a detecção de estados incorretos antes que o defeito seja manifestado; confinamento sendo refletido nas técnicas de isolamento da falha à componente faltosa para limitação da propagação de erros no sistema; recuperação de erros como a transição de um estado de erro para um estado correto; e o tratamento de falhas como a identificação e reparação da falha em si.

A detecção de erros, contextualizando às técnicas utilizadas na DeLIA, pode ser feita por monitoramento de batimentos, onde periodicamente é realizada a troca de mensagens entre componentes para sinalização de seu funcionamento, ou por detecção de sinais de terminação, que podem ser enviados por componentes externas como protocolos de comunicação e escalonadores para sinalizar uma interrupção iminente por parte deles, sendo aproveitável também por ambientes preemptivos \cite{DeLIA_tese}. As limitações desses métodos se dão nas suas capacidades de detecção: o monitoramento por batimentos garante somente o funcionamento da componente para envio dos batimentos, enquanto que sinais de terminação exigem outro componente, geralmente centralizado, sinalizando a interrupção.

O confinamento de erros, enquanto que não tratado pela solução proposta, é resolvido no modelo \textit{fail-stop} pela assumida interrupção imediata da componente relacionada. Fora dele, pode ser tratado com o uso de ações atômicas, isolamento das componentes ou transações, entre outros \cite{weber_ft_conceitos_e_ex}.

A recuperação de erros, em PAD, se dá convencionalmente pela estratégia de pontos de salvamento e recuperação \cite{fault_tolerance_techniques_dongarra}. Baseia-se no salvamento do estado da aplicação (dados necessários para resumo da aplicação de maneira determinística) periodicamente enquanto se está num estado apropriado para, na detecção de um estado de erro, se possa retornar ao estado anterior, assumidamente correto, reduzindo a perda decorrente da falha \cite{survey_fault_tolerance_treaster}. É uma técnica amplamente explorada, com variações quanto a coordenação das componentes para realização do salvamento e decomposições entre o escopo dos dados do estado da aplicação, onde parte do estado pode ser específico à componente (local) ou comum ao sistema a que pertence (global e, neste caso, o sistema sendo a aplicação) \cite{DeLIA_tese}.

No caso específico do salvamento, esforços para se estimar o período ótimo de salvamento numa aplicação resultaram na fórmula de Young/Daly \cite{fault_tolerance_techniques_dongarra}. A fórmula, na versão de \cite{fault_tolerance_techniques_dongarra}, é expressa na equação \eqref{eqYD}, onde $T_{FO}$ é o período ótimo, $\mu$ é o tempo médio entre falhas, $D$ é o tempo decorrido para a aplicação se recuperar da falha, e $C$ é o tempo para realização do salvamento.

\begin{equation}
    T_{FO} = \sqrt{2(\mu - D+R )C}\label{eqYD}
\end{equation}

O tratamento de falhas é uma parte mais especializada, ocorrente ex-post-facto e dependente de contexto \cite{weber_ft_conceitos_e_ex}. Para falhas em nós computacionais em PAD, assumindo devidas etapas de detecção, confinação e recuperação realizadas, pode ser feita pela exclusão do nó defeituoso para resumo da aplicação, uma reinicialização do sistema ou pela solicitação de uma realocação de recursos para o escalonador.

A biblioteca DeLIA, fornece os recursos de detecção de erros e recuperação descritos previamente, com as especificidades de usar o protocolo UDP para uma execução eficiente do monitoramento de batimentos e de salvamentos de dados locais e globais da aplicação. É desenvolvida com foco no modelo de falhas fail-stop e assume que as aplicações sigam o modelo computacional \textit{Bulk Synchronous Parallel} (BSP) \cite{DeLIA_tese}.

O modelo BSP é baseado nas premissas de vários componentes computacionais, conectados entre si por uma componente roteadora que permita a comunicação entre duas componentes, e a capacidade de sincronização entre todas as componentes ou subconjuntos delas a intervalos regulares \cite{bsp_valiant}. A computação da aplicação se dá por etapas globais e locais, onde entre cada etapa global há um conjunto de tarefas repartido entre as componentes para ser executado, com possibilidade de comunicação entre componentes. Quando este conjunto de tarefas é completado, sincroniza-se o estado global da aplicação e segue-se à próxima etapa global. Assumindo este modelo computacional, a DeLIA consegue se aproveitar da sincronização no fim da etapa global para realização de salvamentos do estado global enquanto que, assumindo independência entre tarefas das etapas locais, disponibiliza recursos de salvamento do estado local.

O método de inversão completa da forma de onda em 4D (FWI 4D, do inglês \textit{Full Waveform Inversion} 4D), aplicação do estudo de caso, é uma técnica de imageamento sísmico usada para monitoramento de reservatórios, onde através da aplicação do FWI a dados coletados entre um intervalo de tempo é possível estimar as mudanças no reservatório neste intervalo com base nos modelos obtidos na inversão \cite{fwi4d_obn}. O FWI por si é um algoritmo iterativo de inversão de onda completa baseado na estimativa de um modelo físico onde, para uma entrada previamente determinada, produza um sismograma próximo do sismograma obtido experimentalmente, constituindo um problema de otimização onde se deseja minimizar o erro de aproximação dos sismogramas realizando ajustes no modelo estimado.

\section{Estado da arte}

Como para tolerância a falhas em PAD se tem diversas abordagens, vale a pena discernir o escopo da solução proposta. A primeira divisão se dá entre soluções baseadas em \textit{hardware} ou \textit{software}, de onde soluções em hardware costumam integrar correção e mascaramento de falhas como em memórias RAM com códigos de correção de erro, enquanto que soluções em \textit{software} comumente não necessitam atualizações na infraestrutura dos sistemas computacionais em que são executadas.

Em seguida, soluções baseadas em \textit{software} podem ser elaboradas em diferentes meios: sistema, linguagem, modelo de paralelização, escalonador e aplicação. Enquanto que há certa difusão nessa divisão, dado que aplicações possam pertencer a múltiplas categorias, podemos caracterizar soluções baseadas no sistema ou escalonador como soluções mais gerais, podendo ser invisíveis à aplicação às custas de um menor controle e adequabilidade às especificidades da aplicação; soluções baseadas em linguagem ou modelo de paralelização são aplicadas em ferramentas como linguagens de domínio, compiladores especializados e bibliotecas/protocolos de paralelismo e suas extensões, onde as soluções podem se adequar às aplicações com maior profundidade, mas exigem o desenvolvimento das aplicações em torno delas, criando não só um bloqueio tecnológico na aplicação, como também requisitando uma bagagem técnica em ferramentas específicas a PAD ainda que não sejam consolidadas na área; soluções a nível de aplicação, ao qual nossa proposta pertence, são implementadas através de bibliotecas ou especificamente à aplicação.

Nesta última subdivisão, as soluções tendem a obter a especificidade das soluções às custas da reusabilidade delas, através de bibliotecas que ou se especializam a uma técnica e/ou se comprometem a outras ferramentas da aplicação (assemelhando-se às soluções baseadas em modelo de paralelização) ou a aplicação direta das técnicas no desenvolvimento da aplicação, onde, apesar de permitir técnicas baseadas no próprio algoritmo da aplicação, tendem a ser trabalhadas caso-a-caso, requerendo um maior esforço de desenvolvimento ao longo do tempo.

Para C e C++, linguagens mais convencionais a PAD, as opções conseguem abranger todas essas categorias, como apresentado em \cite{DeLIA_artigo} \cite{survey_fault_tolerance_treaster} \cite{fault_tolerance_techniques_dongarra}. Para Python e Julia, entretanto, existem menos opções. Em Python, existem soluções a nível de modelos de paralelização como PySpark (interface para Apache Spark/Databricks), Parsl \cite{parsl} e SBN-Python \cite{SBNPython}, com o último em específico sendo menos acessível; enquanto que a nível de aplicação existem as implementações específicas, como o salvamento de tensores em bibliotecas de aprendizado profundo como TensorFlow \cite{tensorflow2015-whitepaper} ou a detecção e repetição automática de etapas falhas na análise de genomas na biblioteca xGAP \cite{python_xgap_ftex}, a Fault-Tolerance Interface \cite{FTI}, que implementa salvamento com codificação para códigos de correção de erro, não dispõe de interface em Python no seu repositório público apesar da menção a linguagem em sua implementação. Para Julia, as opções são ainda menores, dispondo somente da capacidade de integrar soluções próprias em modelos de paralelização, como o parâmetro "on\_error" na função "pmap" do módulo Distributed ou compatibilidade com a extensão User-Level Fault Mitigation (ULFM) na interface pra MPI na linguagem \cite{MPI_julia}.

Diante disso, as interfaces ajudam a preencher o vazio de alternativas para implementação de técnicas de tolerância a falhas por bibliotecas de uma forma agnóstica ao protocolo de paralelismo ou biblioteca usada, enquanto dispõe de capacidades de salvamento, detecção de sinais de terminação e monitoramento por batimento que mesmo em C/C++ não se encontram em uma única biblioteca \cite{DeLIA_artigo}.

\section{Metodologia}

Como as interfaces propostas são independentes do estudo de caso, tratamos separadamente cada etapa. Em específico, a seção do estudo de caso aborda a reinterpretação da métrica usada para o custo computacional relativo relacionando-a com a fórmula de Young/Daly mencionada previamente no embasamento teórico.

\subsection{Desenvolvimento das Interfaces}

A possibilidade de fazer tais interfaces parte, primeiramente, da capacidade inerente das linguagens em interagirem com código em C de bibliotecas compartilhadas. Com base nisso, são feitas ferramentas para facilitação desse processo, de onde foi escolhido Cython \cite{cython} para DeLIAP e CxxWrap para DeLIAJ.

Acompanhando a escolha do gerador de interfaces, foram elaborados testes de funcionalidade para as funções exportadas, baseados em resultados esperados em um programa próprio de cada linguagem; que então é usado por um \textit{script} de linha de comando para testar todas as funcionalidades e retornar se o programa passou todos os testes. O \textit{script}, por sua vez, pode ser usado tanto manualmente quanto por ferramentas de integração contínua.

Como a área atuante principal da biblioteca é a de PAD, o programa de teste é paralelizado. Para paralelização dos programas em cada linguagem, foi escolhido o mpi4py \cite{mpi4py_DALSIN} para Python e, para Julia, o módulo Distributed de sua biblioteca padrão, reforçando a independência de uma ferramenta específica de paralelismo para uso da DeLIA.

Por fim, a documentação é feita com Doxygen e sua extensão para Markdown, onde ela complementa a documentação da biblioteca base. O versionamento do código se dá por git, no repositório da DeLIA, na organização do LAPPS dentro da plataforma GitLab, sendo disponível como código aberto.

\subsection{Estudo de Caso}

O estudo de caso é baseado na aplicação da interface desenvolvida para Julia em uma base de códigos de FWI 4D, como citado anteriormente. A aplicação alvo é ativamente usada por pesquisadores no supercomputador do Núcleo de Processamento de Alto Desempenho (NPAD), da UFRN, constituindo em um ambiente realista para validação.

Para desenvolvimento e testagem mais ágil, foi desenvolvido um código exemplo usando a base de códigos para resolver um problema menor de FWI 4D, diminuindo a carga da aplicação. Como a realização de salvamentos, a operação de maior custo computacional intuitivamente, tem complexidade linear, pode-se assumir que os resultados obtidos não tenderão a piorar conforme se aumenta a escala de dados — ao menos até que se chegue a gargalos de escrita em alguma das memórias. O processo de integração da interface ao código é disposto antes da análise de \textit{overhead}, complementando o relato.

Para a análise quantitativa foram feitas duas versões da aplicação, uma sem a DeLIAJ e outra com ela, e foram feitas dez execuções em cada, com 20 iterações do algoritmo em um conjunto de dados com 50 amostras, distribuídos em 32 núcleos de um único nó da partição Intel-512 do NPAD, usado para as duas medições. O número baixo de execuções totais se deu pelo tempo exigido para coleta dos dados, mas pelo custo adicional surgir nos momentos de chamada da interface, que é, novamente, linear com o número de iterações, então o custo computacional relativo de cada execução é composto do custo computacional de cada iteração global da aplicação, amortizando efeitos de \textit{outliers}, junto a um fator constante de preparação dos recursos da biblioteca. Foram usadas somente detecção de sinais de terminação e salvamento global na aplicação por questões a serem apresentadas junto aos resultados.

A métrica usada para o custo computacional foi o tempo de execução do nó principal, sendo utilizada para gráficos de \textit{box-plot} e cálculo do custo computacional mediano relativo, modificado a partir do usado no artigo da DeLIA \cite{DeLIA_artigo} (equação \eqref{eqCCMR}) , e um gráfico de dispersão para visualização dos tempos individuais de cada execução. Enquanto os gráficos ajudam a ver as próprias distribuições de tempo obtidas, o uso do custo computacional mediano relativo se dá pela sua maior resistência a valores extremos em relação à média. Diferentemente do artigo da biblioteca base, não é feita comparação com o desvio padrão relativo por, nestes experimentos, não se fazer o ajuste fino da estratégia de salvamento, realizando o salvamento a cada iteração global — que, como apresentado no embasamento teórico, não necessariamente é uma estratégia ótima. Na equação a seguir, $M_{DeLIAJ}$ e $M_{original}$ são as medianas do tempo de execução das versões com e sem a DeLIAJ, respectivamente.

\begin{equation}
    overhead = \frac{M_{DeLIAJ} - M_{original}}{M_{DeLIAJ}}\label{eqCCMR}
\end{equation}

Podemos estabelecer uma relação entre o custo computacional mediano relativo obtidos e o desenvolvimento da variante da fórmula de Young/Daly para o período ótimo de salvamento apresentada em \cite{fault_tolerance_techniques_dongarra}. As condições do experimento permitem a estimativa da componente do custo relativa a execução sem falhas, expressa por \eqref{eqWFF1}, onde $W_{FF}$ é o custo computacional sem falhas, $T_{FF}$ é o tempo de execução total sem falhas com a biblioteca e $T_{base}$ é o tempo de execução total sem falhas sem ela. Como se realiza o salvamento em todas as iterações globais, o período de salvamento é mínimo e, consequentemente, estimamos o custo computacional relativo máximo para esta componente.

\begin{equation}
    W_{FF} = \frac{T_{FF} - T_{base}}{T_{FF}}\label{eqWFF1}
\end{equation}

A aquisição e o tratamento dos dados levaram aproximadamente três dias no total, sendo feitos gráficos de \textit{box-plot} e dispersão para visualização da distribuição de tempos obtidos, com o último sendo viabilizado pelo baixo número de amostras totais.

\section{Resultados e Discussão}

Atualmente, o código fonte das interfaces está disponível publicamente no repositório da DeLIA. O desenvolvimento e atualização das interfaces às mudanças na biblioteca principal não são sincronizados, então eventuais incompatibilidades com versões mais recentes são possíveis, como em uma recente reestruturação na estrutura de arquivos da DeLIA. Casos semelhantes podem se dar com as dependências das interfaces, como ocorrido durante o desenvolvimento da interface em Julia e o CxxWrap.

A interface disponibiliza as capacidades mencionadas no artigo original da DeLIA \cite{DeLIA_artigo}, com pendências para integração das capacidades de \textit{failover} e replicação introduzidas mais recentemente \cite{DeLIA_tese}. Por limitações de Julia quanto ao uso de classes, DeLIAJ não disponibiliza as classes abstratas para salvamento, usando para isso unicamente as funções que inscrevem ponteiros para salvamento. Apesar destes contratempos, tanto a DeLIAP quanto DeLIAJ tiveram sucesso nos seus testes de funcionalidade automatizados e aplicações-exemplo — versões paralelas de algoritmos de busca simples baseados em meta-heurística, nomeadamente Enxame de Partículas e Evolução Diferencial. Pelo curto tempo de execução destes exemplos, não valeria fazer análise do custo computacional destas aplicações.

A aplicação do estudo de caso também teve funcionamento adequado, mas especificidades na forma que o paralelismo era empregado na aplicação inviabilizavam o uso de salvamentos locais enquanto que, pela falta de atomicidade das funções da base de código, não se podem aplicar corretamente o monitoramento de batimentos ou a detecção de sinais de terminação. O primeiro problema se trata da forma como as iterações locais eram atribuídas, onde os parâmetros das funções executadas, incluindo o que deveriam ser os dados locais, eram passados diretamente do nó principal aos demais nós, sem alternativa prática para armazenamento do estado local no ponteiro configurado para esse; para o outro problema, por sua vez, tem-se a perspectiva que possa ser desenvolvido, para a biblioteca base, uma solução para tornar funções "atômicas" para os fins de salvamento da biblioteca, diminuindo os pré-requisitos para aplicação das técnicas de detecção de erro.

\begin{figure}[htbp]
\centerline{\includegraphics[scale=0.5]{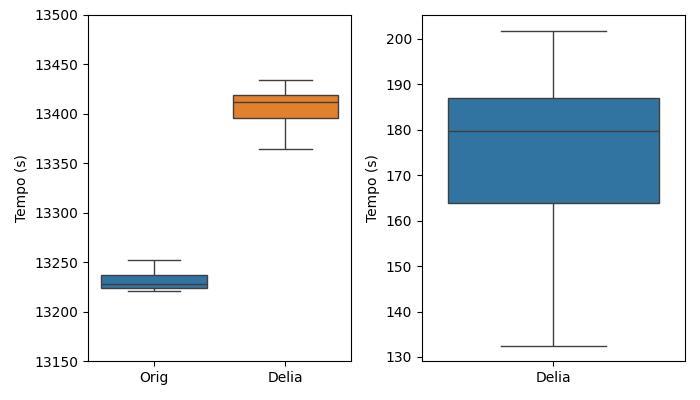}}
\caption{\textit{Box-plots} das amostras coletadas.}
\label{figBoxplots}
\end{figure}

Em questão de desempenho, o custo computacional relativo estimado ficou em torno de 1,4\%, com valor absoluto de aproximadamente 174,9448 segundos para os 13441,8312 segundos medianos da execução total com a biblioteca. Pelo gráfico na figura \ref{figBoxplots} é possível ver à esquerda os dois \textit{box-plots} das amostras em uma mesma escala para comparação, onde, para uma escala duas ordem de grandeza menor que o tempo de execução, podem-se considerar próximos. À direita outro \textit{box-plot}, baseado na diferença entre cada par de amostras, tenta refletir a mesma ideia, apesar de sugerir uma temporalidade que não é assumida no experimento.

\begin{figure}[htbp]
\centerline{\includegraphics[scale=0.35]{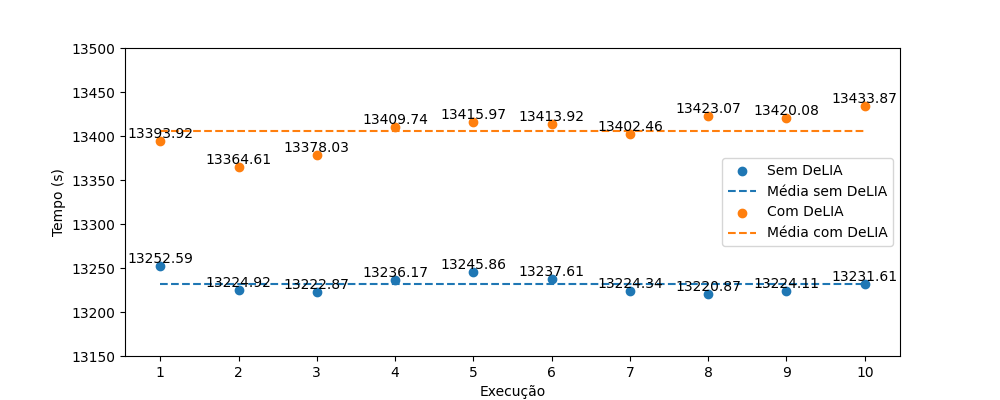}}
\caption{Gráfico de dispersão das amostras coletadas.}
\label{figScatter}
\end{figure}

Um olhar mais profundo sobre as amostras pode ser feito na figura \ref{figScatter}. Apesar de flutuações no desempenho afetarem as duas versões, a versão com a DeLIAJ foi mais impactada por elas, refletindo-se em um desvio-padrão quase duas vezes maior que o original (21.3290 e 10.7677 cada). Isso pode ser indicativo de um envolvimento da escrita na memória secundária com as flutuações, mas as diferentes possibilidades — sistema de arquivos distribuído, processador, SSD —, aliados à falta de amostras, não permitem mais que especulações sobre a origem e/ou localidade do fenômeno.

Uma versão do código paralelizada em MPI, através da biblioteca MPI.jl \cite{MPI_julia}, também foi desenvolvida, dessa vez com capacidades para salvamento local. Tal mudança, entretanto, exigiria a reexecução de testes para a versão base com MPI e com a DeLIA, além de avaliações entre a viabilidade de uso do salvamento local onde, pela escassez de tempo, não poderão ser apresentados aqui.

\section{Conclusão}

Foram propostas e desenvolvidas as interfaces da DeLIA, biblioteca de confiabilidade para aplicações iterativas, para Julia e Python a fim de tornar tolerância a falhas mais acessível em PAD com a extensão das capacidades da biblioteca base para linguagens mais fáceis e ainda bem estabelecidas em computação científica como um todo.

Cada funcionalidade mencionada no artigo da DeLIA \cite{DeLIA_artigo}, exceto onde limitado pelas linguagens-alvo, estão disponíveis, sendo verificadas por testes automatizados. Atualizações para inclusão de novas capacidades introduzidas na biblioteca base e manutenção da compatibilidade com essa e outras dependências se fazem um processo contínuo, realizado colaborativamente junto ao repositório da DeLIA, de código aberto.

Buscou-se validar a eficácia das interfaces para prover tolerância a falhas em aplicações PAD através da integração da DeLIAJ em uma base de códigos do método de inversão completa da forma de onda em 4D em Julia. Nos testes, realizados em uma versão reduzida do problema através da comparação de tempos de execução com e sem o uso da interface, obteve-se um custo computacional relativo de aproximadamente 1,4\%, favorável à eficiência computacional da interface. Como contraponto, fatores como a forma que o módulo de paralelismo escolhido foi usado inviabilizaram a aplicação de salvamentos globais, enquanto que a falta de atomicidade nas funções da base de código impossibilitavam o uso de detecção de sinais ou detecção de batimentos, expondo um ponto negativo no primeiro caso e uma oportunidade de melhoria da biblioteca para o último. Consideramos as interfaces viáveis para utilização externa, apesar das reconhecíveis limitações a serem resolvidas.

\section*{Reconhecimentos}
Agradecemos ao Núcleo de Processamento de Alto Desempenho pela disponibilização do supercomputador e ao projeto "Novas metodologias computacionalmente escaláveis para sísmica 4D orientado ao alvo em reservatórios do pré-sal", que tornou este trabalho possível.

\bibliography{referencias}
\bibliographystyle{ieeetr}

\end{document}